\documentclass{ptapap}

\usepackage{comment}

\author{Pavlo Kashko}[UoM,TSU]
\author{Viktor Khalack}[UoM]
\author{Oleksandr Kobzar}[UoM]
\author{Dmytro Tvardovskyi}[UoM,ODM]
\author{Mathieu Perron-Cormier}[UoM]
\affil[UoM]{D\'epartement de Physique et d'Astronomie, Universit\'e de Moncton, Moncton, N.B., Canada E1A 3E9}
\affil[TSU]{Faculty of Physics, Taras Shevchenko National University of Kyiv, Kyiv, Ukraine}
\affil[ODM]{Faculty of Mathematics, Physics and Information Technology, Odessa National I.I. Mechnikov University, Odessa, Ukraine }

\title{Revealing the nature of HD~63401}

\begin{document}

\maketitle

\begin{abstract}

HD~63401 is known magnetic chemically peculiar (mCP) star that shows a slow rotation and probably possesses a hydrodynamically stable stellar atmosphere. In the latter case the atomic diffusion mechanism enforced by the magnetic field can lead to stratification of elemental abundance with optical depth. HD~63401 was recently observed with the space telescope \textit{TESS}, and its light curve shows distinct variability, which is usually detected in the $\alpha^2$ CVn type stars. Based on the analysis of the light curve we derived its rotational period $P$ = 2.414 $\pm$ 0.018~d and studied variability of the effective temperature with this period. The estimates of the effective temperature and surface gravity were obtained from the best fit of Balmer line profiles observed in seven high-resolution spectra acquired with the spectropolarimeter ESPaDOnS for HD~63401. The same spectra were used to perform an abundance analysis employing the modified ZEEMAN2 code. We have found that He, C, P, V, Y, and Dy are in deficit in stellar atmosphere of HD~63401, while Na, Al, Si, Fe, Zn, and Sr are significantly overabundant.

\end{abstract}

\section{Introduction}

HD~63401 is a relative bright chemically peculiar (CP) star of spectral type B9 \citep{Renson+91} with significant magnetic field \citep{Bagnulo+06,Kochukhov+Bagnulo06}.
\citet{Bailey+14} estimated its global stellar parameters and age $\log{t} = (7.7\pm0.1)$ considering the fact that HD~63401 is a member of the open cluster NGC~2451. These authors have also detected the enhanced abundance of Si, Ti, Cr, Fe, and Pr in stellar atmosphere of HD~63401, while He, O and Mg seem to be in deficit. \citet{Bailey+14} concluded that the abundance peculiarities observed for different chemical elements may change with the stellar evolution on the main sequence.

Based on the analysis of the Hipparcos photometry data \citet{Adelman+00} reported a detection of photometric variability of HD~63401 with amplitude 60$\pm$2 mmag. By analysing uvby-photometry \citet{Hensberge+76} derived a rotation period of HD~63401 as $P$ = 2.41 $\pm$ 0.02~d.
This star has been recently observed with the Transiting Exoplanet Survey Sattelite (\textit{TESS}) \citep{Ricker+15} that provided short cadence (2~min) photometric measurements.
HD~63401 was included to the list of relatively bright CP stars that show photometric variability related to the stellar rotation \citep{Kobzar+19}. This star (TIC~175604551) was observed in the sectors 7 and 8.

\section{Analysis of the TESS data}

\begin{figure}[h!]
  \centering
  \begin{minipage}[t]{0.48\textwidth}
    \includegraphics[width=\textwidth]{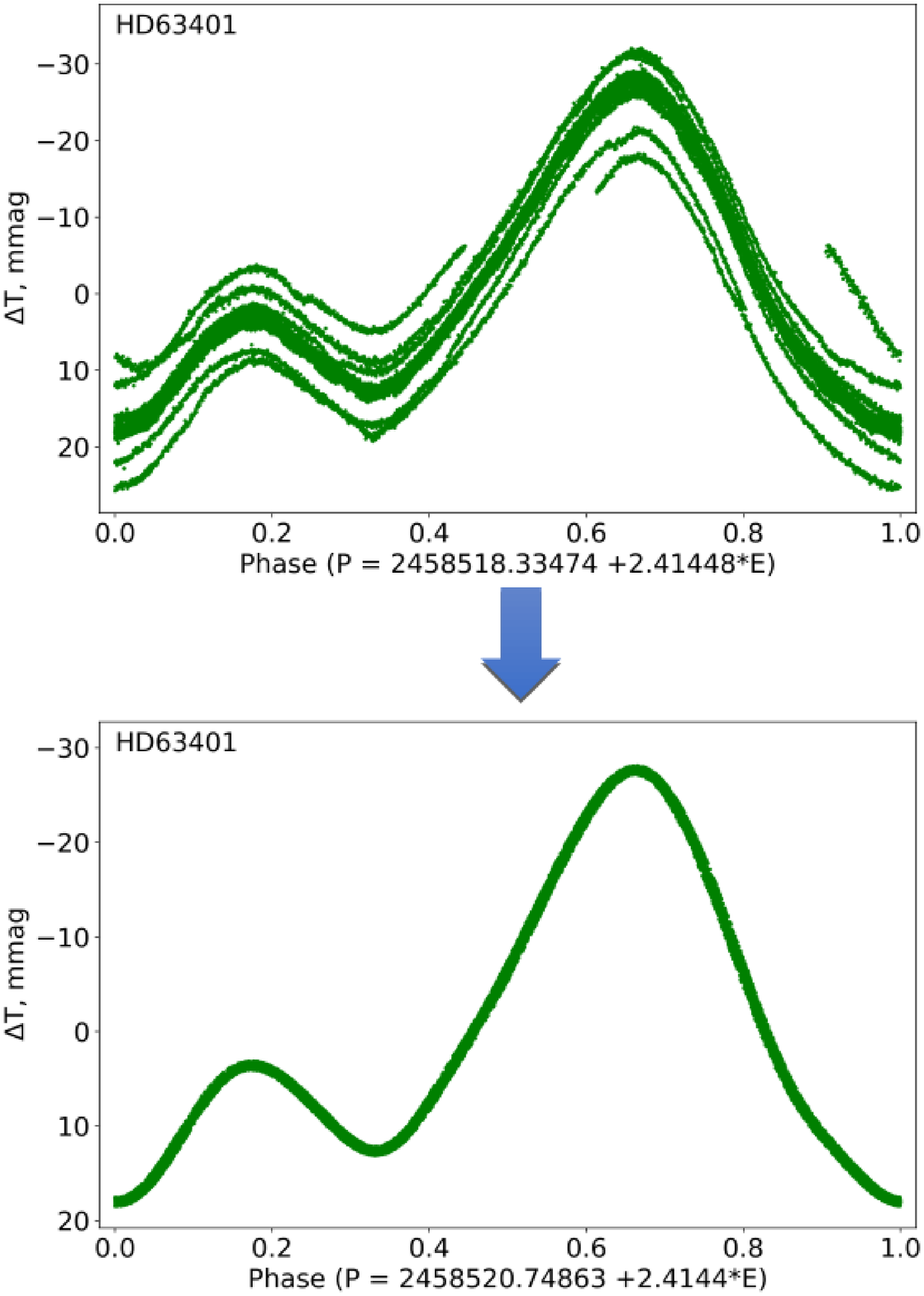}
    \caption{The \textit{TESS} photometric data phased with the period derived from the Fourier analysis (upper panel) and the reduced LC phase diagram (lower panel) produced by the code Period~D\&P for HD~63401 using the refined period.}
    \label{fig:Cutting}
  \end{minipage}
  \quad
 \begin{minipage}[t]{0.48\textwidth}
    \includegraphics[width=\textwidth]{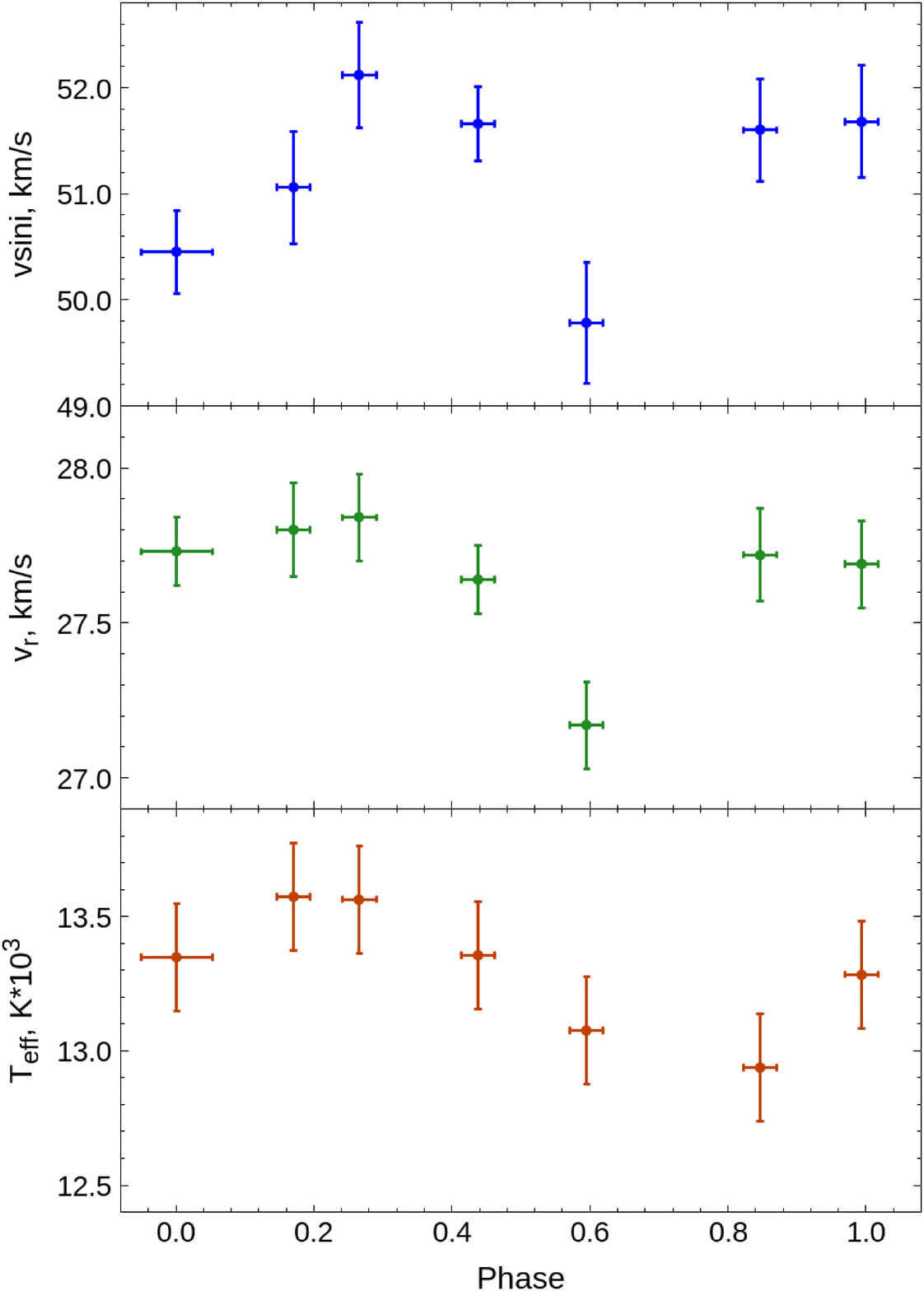}
    \caption{The v$\sin{i}$, v$_{\rm r}$, and  T$_{\rm eff}$ were obtained from the spectral analysis. Their rotational phases were calculated according to the derived period.
    }
    \label{fig:Speeds}
  \end{minipage}
\end{figure}

The photometric data obtained with \textit{TESS} for HD~63401 were downloaded via the Mikulski Archive for Space Telesopes\footnote{https://archive.stsci.edu/tess/bulk\_downloads.html} (MAST) and are publicly available.
To perform the analysis of its light curve an automatic procedure TESS-AP \citep{Khalack+19a} was used to estimate the frequencies, amplitudes and phases of analyzed signals.
The procedure TESS-AP consists of several scripts and codes including the code Period04 \citep{Period04} and was designed for automatic data analysis.
The code Period~D\&P\footnote{GitHub: https://pawakawko.github.io/Period-D-P} was developed using the Python language and employed to refine the value of stellar rotation period.
The code uses photometric observations to construct a phase curve, and fits it with the next Fourier series:
\begin{equation}
    m_j = A_0 + \sum_{n=1}^{N} A_n \cdot \cos(\frac{2\pi n \cdot t_j }{P} + \phi_n),
\end{equation}
where $m_j$ is an observed magnitude, $t_j$ - corresponding time of observation, N - number of harmonics, P - rotation period, $A_n$ and $\phi_n$ - amplitudes and phases of the harmonic series. To derive the best fit parameters we used the method of non-linear least squares approximation \citep{marquardt:1963}. The code Period~D\&P calculates a value of the rotation period and its precision and plots phase curve. The light curve may contain trends, flares and other features. The trends mainly occur at the beginning or at the end of each continuous sequence of the \textit{TESS} observations and do not repeat itself with the rotation period. 
To exclude a contribution of the trends the code Period~D\&P cuts off all measurements that have too large deviation from the best fit and repeats the fitting procedure (see Fig.~\ref{fig:Cutting}). This procedure resulted in the estimation of the stellar rotation period of $P$ = 2.414 $\pm$ 0.018~d for HD~63401.

\section{Spectral analysis}

We have performed a spectral analysis of seven high-resolution ($R = 65000$) and high signal-to-noise ratio Stokes IV spectra recently obtained with the spectropolarimeter ESPaDOnS at the CFHT. The dedicated software package Libre-ESpRIT \citep{Donati+97} was employed to reduce the obtained data. The non-normalized spectra were used to fit the Balmer lines observed in Stokes I spectra by the theoretical profiles with the help of FITSB2 code \citep{Napiwotzki+04} employing the grids of stellar atmosphere models \citep{Khalack+LeBlanc15a}. The best fit values of the $T_{\rm eff}$ derived for different rotational phases from the analyses of each spectrum are shown in Fig.~\ref{fig:Speeds} (see lower panel).

The Stokes IV spectra were normalized to conduct abundance analysis with the help of ZEEMAN2 code \citep{Landstreet88}. An automatic procedure \citep{Khalack18} was used to select blended and unblended line profiles suitable for abundance analysis. Model of stellar atmosphere was calculated for the derived average parameters $T_{\rm eff} = 13360\pm200~K$, $\log{g} = 4.1 \pm0.2$, and metallicity $M = 0$ with the assistance of PHOENIX-15 code \citep{Hauschildt+97}. This model was employed by the code ZEEMAN2 to fit the selected line profiles with the synthetic ones. The average abundance of chemical elements in stellar atmosphere of HD~63401 was estimated for each rotational phase as well as the values of v$\sin{i}$ and radial velocity v$_{\rm r}$ (see upper two panels in Fig.~\ref{fig:Speeds}). The abundance estimates averaged over seven phases for each studied chemical element
are presented in Table~\ref{tab:Abundance_short}

\begin{table}[t]
	\centering
	\caption{Average abundance 
calculated for the studied chemical species from the analysis of all seven spectra of HD~63401.}
	\label{tab:Abundance_short}
	\begin{tabular}{lc|lc|lc|lc}
		\hline
	El. & [X/H] & El. & [X/H] & El. & [X/H] & El. & [X/H] \\
		\hline
	He &  -1.25 $\pm$ 0.28 & P  & -1.54 $\pm$ 0.71 & Ba &   0.44  $\pm$ 0.28 & Ce &  -0.39  $\pm$ 0.08 \\
	C  &  -0.94 $\pm$ 0.21 & Ca & -0.20 $\pm$ 0.16 & Co &   1.38  $\pm$ 0.95 & Pr &   0.72  $\pm$ 0.28 \\
    N  &  -0.09 $\pm$ 0.18 & Sc &  0.42 $\pm$ 0.08 & Ni &   0.08  $\pm$ 0.37 & Nd &   0.16  $\pm$ 0.17    \\
    O  &  -0.31 $\pm$ 0.19 & Ti &  0.44 $\pm$ 0.12 & Cu &  -0.19  $\pm$ 0.04 & Sm &  -0.37  $\pm$ 0.06 \\
    Ne &  -0.24 $\pm$ 0.03 & V  & -1.75 $\pm$ 0.05 & Zn &   2.60  $\pm$ 0.17 & Gd &  -0.05  $\pm$ 0.13 \\
    Na &   1.63 $\pm$ 0.48 & Cr &  0.83 $\pm$ 0.17 & Sr &   1.16  $\pm$ 0.45 & Dy &  -0.72  $\pm$ 0.52 \\
    Al &   1.74 $\pm$ 0.65 & Mn &  0.55 $\pm$ 0.17 & Y  &  -0.64  $\pm$ 0.06 & \\
    Si &   1.07 $\pm$ 0.61 & Fe &  1.57 $\pm$ 0.18 & Zr &  -0.28  $\pm$ 0.10 & \\
		\hline
	\end{tabular}
\end{table}

\begin{figure}[h]
\centering
\includegraphics[width=0.95\textwidth]{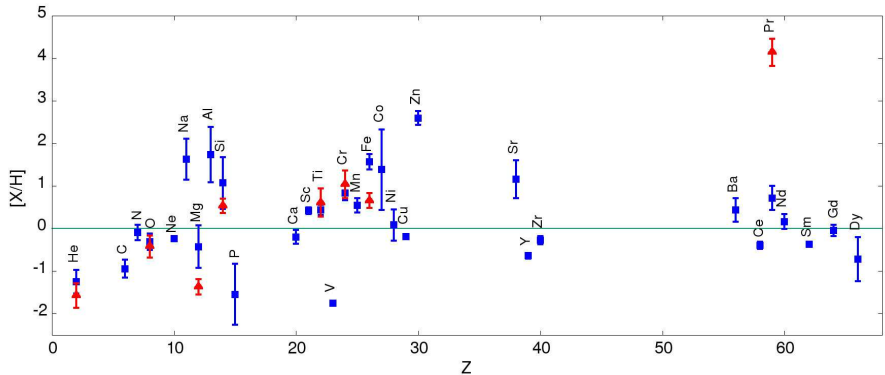}
\label{fig:Abundance}
\caption{Comparison of the derived average abundances of chemical elements (squares) in stellar atmosphere of HD~63401 with the data (triangles) reported by \citet{Bailey+14}. The green line represents the solar abundance \citep{Grevesse+15}.}
\end{figure}


\section{Discussion}

We have used the photometric data provided by \textit{TESS} for HD~63401 to derive its rotation period of $P$ = 2.414 $\pm$ 0.018~d considering that a presence of overabundance patches causes redistribution of the emitted flux forming brighter spots in stellar atmosphere of this star. In the context of oblique rotator model \citep{Stibbs50}, the overabundance patches and brightness spots will change their visibility with rotational phase and lead to the observed flux variability (see \citealt{David-Uraz+19}). This hypothesis is in a good accordance with the detected weak variability of the derived effective temperature with rotational phase. It may be noticed that the maxima of the LC phase diagram (lower panel in Fig.~\ref{fig:Cutting}) have similar rotational phases as the extrema at the temperature variability (lower panel in Fig.~\ref{fig:Speeds}). However, the minimum of the effective temperature refers almost to the maximum of the LC. The average values of v$\sin{i}$ and radial velocity derived from the analysis of over hundred line profiles in each studied spectrum also show a weak variability with rotational phase similar to the one detected for $T_{\rm eff}$ (see Fig.~\ref{fig:Speeds}).

Our estimates of average (over seven phases) abundances of chemical elements are in a relatively good accordance with the data published by \citet{Bailey+14} (see Fig.~\ref{fig:Abundance} and Table~\ref{tab:Abundance_short}). We have found that He, C, P, V, Y, and Dy are significantly underabundant in stellar atmosphere of HD~63401, while Na, Al, Si, Fe, Zn, and Sr show strong overabundance. Our estimate of iron abundance appears to be significantly higher than the one reported by \citet{Bailey+14}.
The abundance of praseodymium derived in this work is much lower that the one obtained by \citet{Bailey+14}.
All analyzed rare earth elements, meanwhile, show abundance close to their solar abundance \citep{Grevesse+15}.

The detected variability of the light curve and $T_{\rm eff}$, and the strong overabundance of some chemical elements support the hypothesis of abundance patches existence in stellar atmosphere of HD~63401. These patches can be formed and supported by the effective mechanism of atomic diffusion \citep{Michaud70}. Therefore, we consider that stellar atmosphere may be hydrodynamically stable and plan to search for abundance stratification of chemical elements with optical depth in this star.

\acknowledgements{P.K and D.T are thankful to the Mitacs Globalink Research Internship program for the support of this research. V.K. and M.P.-C. acknowledge the support from the Natural Sciences and Engineering Research Council of Canada. M.P.-C., O.K. and V.K. are thankful to the Facult\'{e} des \'{E}tudes Sup\'{e}rieures et de la Recherche and to the Facult\'{e} des Sciences de l'Universit\'{e} de Moncton for the financial support of this research. Part of calculations was done on the supercomputer \textit{b\'{e}luga} of the \'{E}cole de technologie sup\'{e}rieure in Montreal, under the guidance of Calcul Qu\'{e}bec and Calcul Canada. Analysed spectra were obtained at the Canada-France-Hawaii Telescope (CFHT) which is operated by the National Research Council of Canada, the Institut National des Sciences de l'Univers of the Centre National de la Recherche Scientifique of France, and the University of Hawaii. The operations at the Canada-France-Hawaii Telescope were conducted with care and respect from the summit of Maunakea which is a significant cultural and historic site.
This paper includes data collected by the \textit{TESS} mission. Funding for the \textit{TESS} mission is provided by the NASA Explorer Program. This research used the SIMBAD database, operated at CDS, Strasbourg, France. Photometric data presented in this paper were obtained from the Mikulski Archive for Space Telescopes (MAST).}

\bibliographystyle{ptapap}
\bibliography{Kashko}

\begin{thebibliography}{21}
\providecommand{\natexlab}[1]{#1}
\providecommand{\url}[1]{\texttt{#1}}
\providecommand{\urlprefix}{URL }
\providecommand{\eprint}[2][]{\url{#2}}

\bibitem[{{Adelman} et~al.(2000){Adelman}, {Gentry}, \& {Sudiana}}]{Adelman+00}
{Adelman}, S.~J., {Gentry}, M.~L., {Sudiana}, I.~M., \emph{Information Bulletin
  on Variable Stars} \textbf{4968}, 1 (2000)

\bibitem[{{Bagnulo} et~al.(2006)}]{Bagnulo+06}
{Bagnulo}, S., et~al., \emph{\aap} \textbf{450}, 777 (2006)

\bibitem[{{Bailey} et~al.(2014){Bailey}, {Landstreet}, \&
  {Bagnulo}}]{Bailey+14}
{Bailey}, J.~D., {Landstreet}, J.~D., {Bagnulo}, S., \emph{\aap} \textbf{561},
  A147 (2014)

\bibitem[{{David-Uraz} et~al.(2019){David-Uraz}, {Neiner}, {Sikora}, \&
  {MOBSTER Collaboration}}]{David-Uraz+19}
{David-Uraz}, A., {Neiner}, C., {Sikora}, J., {MOBSTER Collaboration},
  \emph{this Proceedings}  (2019)

\bibitem[{{Donati} et~al.(1997)}]{Donati+97}
{Donati}, J.~F., et~al., \emph{\mnras} \textbf{291}, 658 (1997)

\bibitem[{{Grevesse} et~al.(2015){Grevesse}, {Scott}, {Asplund}, \&
  {Sauval}}]{Grevesse+15}
{Grevesse}, N., {Scott}, P., {Asplund}, M., {Sauval}, A.~J., \emph{\aap}
  \textbf{573}, A27 (2015)

\bibitem[{{Hauschildt} et~al.(1997){Hauschildt}, {Baron}, \&
  {Allard}}]{Hauschildt+97}
{Hauschildt}, P.~H., {Baron}, E., {Allard}, F., \emph{\apj} \textbf{483}, 390
  (1997)

\bibitem[{{Hensberge} et~al.(1976){Hensberge}, {De Loore}, {Zuiderwijk}, \&
  {Hammerschlag-Hensberge}}]{Hensberge+76}
{Hensberge}, H., {De Loore}, C., {Zuiderwijk}, E.~J., {Hammerschlag-Hensberge},
  G., \emph{\aap} \textbf{48}, 383 (1976)

\bibitem[{{Khalack}(2018)}]{Khalack18}
{Khalack}, V., \emph{\mnras} \textbf{477}, 882 (2018)

\bibitem[{{Khalack} \& {LeBlanc}(2015)}]{Khalack+LeBlanc15a}
{Khalack}, V., {LeBlanc}, F., \emph{\aj} \textbf{150}, 2 (2015)

\bibitem[{{Khalack} et~al.(2019)}]{Khalack+19a}
{Khalack}, V., et~al., \emph{\mnras} \textbf{490}, 2102 (2019)

\bibitem[{{Kobzar} et~al.(2019)}]{Kobzar+19}
{Kobzar}, O., et~al., \emph{this Proceedings}  (2019)

\bibitem[{{Kochukhov} \& {Bagnulo}(2006)}]{Kochukhov+Bagnulo06}
{Kochukhov}, O., {Bagnulo}, S., \emph{\aap} \textbf{450}, 763 (2006)

\bibitem[{{Landstreet}(1988)}]{Landstreet88}
{Landstreet}, J.~D., \emph{\apj} \textbf{326}, 967 (1988)

\bibitem[{{Lenz} \& {Breger}(2005)}]{Period04}
{Lenz}, P., {Breger}, M., \emph{Communications in Asteroseismology}
  \textbf{146}, 53 (2005)

\bibitem[{Marquardt(1963)}]{marquardt:1963}
Marquardt, D.~W., \emph{SIAM Journal on Applied Mathematics} \textbf{11}, 431
  (1963)

\bibitem[{{Michaud}(1970)}]{Michaud70}
{Michaud}, G., \emph{\apj} \textbf{160}, 641 (1970)

\bibitem[{{Napiwotzki} et~al.(2004)}]{Napiwotzki+04}
{Napiwotzki}, R., et~al., {Double degenerates and progenitors of supernovae
  type Ia}, \emph{Astronomical Society of the Pacific Conference Series},
  volume 318, 402--410 (2004)

\bibitem[{{Renson} et~al.(1991){Renson}, {Gerbaldi}, \& {Catalano}}]{Renson+91}
{Renson}, P., {Gerbaldi}, M., {Catalano}, F.~A., \emph{\aaps} \textbf{89}, 429
  (1991)

\bibitem[{{Ricker} et~al.(2015)}]{Ricker+15}
{Ricker}, G.~R., et~al., \emph{Journal of Astronomical Telescopes, Instruments,
  and Systems} \textbf{1}, 014003 (2015)

\bibitem[{{Stibbs}(1950)}]{Stibbs50}
{Stibbs}, D.~W.~N., \emph{\nat} \textbf{165}, 195 (1950)

\end{thebibliography}

\end{document}